\documentclass[12pt,letterpaper,copyedit]{article}
\usepackage{osajnl}

\def\la{\mathrel{\mathchoice {\vcenter{\offinterlineskip\halign{\hfil
$\displaystyle##$\hfil\cr<\cr\sim\cr}}}
{\vcenter{\offinterlineskip\halign{\hfil$\textstyle##$\hfil\cr
<\cr\sim\cr}}}
{\vcenter{\offinterlineskip\halign{\hfil$\scriptstyle##$\hfil\cr
<\cr\sim\cr}}}
{\vcenter{\offinterlineskip\halign{\hfil$\scriptscriptstyle##$\hfil\cr
<\cr\sim\cr}}}}}
\def\ga{\mathrel{\mathchoice {\vcenter{\offinterlineskip\halign{\hfil
$\displaystyle##$\hfil\cr>\cr\sim\cr}}}
{\vcenter{\offinterlineskip\halign{\hfil$\textstyle##$\hfil\cr
>\cr\sim\cr}}}
{\vcenter{\offinterlineskip\halign{\hfil$\scriptstyle##$\hfil\cr
>\cr\sim\cr}}}
{\vcenter{\offinterlineskip\halign{\hfil$\scriptscriptstyle##$\hfil\cr
>\cr\sim\cr}}}}}

\begin{document}

\title{Ray-tracing and physical-optics analysis of the aperture efficiency 
in a radio telescope}

\author{Luca Olmi$^{1,2}$ and Pietro Bolli$^{3,4}$}
\affiliation{
      $^{1}$INAF, Istituto di Radioastronomia, sezione di Firenze,
      Largo E. Fermi 5,
      I-50125 Firenze, Italy}
\affiliation{
      $^{2}$University of Puerto Rico, Rio Piedras Campus,
      Physics Dept., Box 23343, UPR Station, San Juan,
      PR 00931-3343, USA}
\affiliation{
      $^{3}$INAF, Osservatorio Astronomico di Cagliari, 
      Loc. Poggio dei Pini, Strada 54, I-09012, Cagliari, Italy}
\affiliation{
      $^{4}$INAF, Istituto di Radioastronomia, sezione di Bologna, 
      Via P. Gobetti 101,
      I-40129 Bologna, Italy}

%
\email{olmi@arcetri.astro.it}

\begin{abstract}The performance of telescope systems working at microwave 
or visible/IR wavelengths is typically described in terms 
of different parameters according to the wavelength range.
Most commercial ray tracing packages
have been specifically designed for use with visible/IR
systems and thus, though very flexible and sophisticated, do not provide 
the appropriate parameters to fully describe microwave antennas, and 
thus to compare with specifications.
In this work we demonstrate that the Strehl ratio is equal to the 
phase efficiency when the apodization factor is taken into account.
The phase efficiency is the most critical contribution to the
aperture efficiency of an antenna, and the most difficult parameter to
optimize during the telescope design. The equivalence between
the Strehl ratio and the phase efficiency gives the designer/user
of the telescope the opportunity to use the faster 
commercial ray-tracing software to optimize the design.
We also discuss the results of several tests
performed to check the validity of this relationship that we carried out
using a ray-tracing software, ZEMAX and a full Physical Optics
software, GRASP9.3, applied to three different telescope designs that
span a factor of $\simeq 10$ in terms of $D/\lambda$. The maximum
measured discrepancy between phase efficiency and Strehl ratio
varies between $\simeq 0.4$\% and 1.9\% up to an offset angle of 
$>40$ beams, depending on the 
optical configuration, but it is always less than 0.5\% where the
Strehl ratio is $>0.95$.
\\

\end{abstract}


\maketitle 

\section{Introduction}
\label{sec:intro}
Performance evaluation is a critical step in the
design of any optical system, either at microwave or visible/IR wavelengths.
The image quality criteria more commonly used, however, are quite different
in these two regions of the electromagnetic spectrum. In fact, in the analysis
of microwave antennas and radio telescopes the two fundamentals figures-of-merit
used by designers and users are the aperture efficiency and the beam
efficiency, whereas in optical systems the Strehl ratio and ray aberrations are
often quoted.
This is because of the {\it coherent} nature of most microwave antennas,
where single-moded receivers are generally used (exceptions may be millimeter
and submillimeter bolometers used in radio astronomy), making the phase
distribution in the image as important as the amplitude distribution in
determining the performance of the optics. In fact, the aperture efficiency is
intrinsically dependent on the phase distributions since it is calculated as
a correlation integral between the focal region field produced by an incident plane wave and the horn
aperture field.

The difference between the microwave and visible/IR wavelengths regimes, in 
terms of the image quality criteria applied to astronomical telescopes,  
has been reduced over the past 10-15 years thanks to the development of 
focal plane arrays (FPA, hereafter). In fact, the noise performance of 
receivers used in radio astronomy  has improved dramatically during 
this time, especially at millimeter and 
submillimeter wavelengths. As a consequence, it has become clear that
the best means of increasing observing efficiency for mapping extended  
sources or to conduct blind searches 
is to use imaging arrays located at the focal plane of the telescope. 
This implies the need of a larger field of view (FOV) with few aberrations 
in the range of frequencies used by the array(s) of receivers. Very often 
these FPA require some relay optics to convert
the telescope focal ratio (which, in some cases, may be quite large, i.e. 
$\ga 10$) to the smaller focal ratios of the individual feed-horns.
As a consequence, the overall image quality of the {\it total} system, 
telescope and reimaging optics, must be evaluated over a wide FOV,
thus effectively contributing to bridging the gap between the microwave 
and visible wavelengths regimes.

A number of commercial ray tracing packages exist that are being used to 
analyse the performance of FPAs for use with existing or planned
(sub)millimeter telescopes. However, many of these packages have been
specifically designed for use with optical (i.e., visible and IR) 
systems and thus, although
very flexible and sophisticated, they do not provide the appropriate 
parameters to fully describe microwave antennas, and thus to compare 
with specifications.
The possibility to easily convert an optical-based design parameter,
such as the Strehl ratio, to a fundamental antenna-based design
parameter, such as the phase efficiency, gives the designer/user
of the telescope the opportunity to use the faster
commercial ray-tracing software to optimize the design. Once the
design is optimized, a full Physical Optics software can be used
to analyse more thoroughly all critical performance parameters
of the antenna (e.g., spillover, antenna noise temperature, etc.).
Another advantage offered by this conversion consists of the 
possibility to study the degrading effects on the wavefront caused
by obstructions to the beam (e.g., secondary reflector and its
support struts) which are notoriously difficult to simulate in
Physical Optics software.

In this paper we review the main design parameters generally used in 
evaluating the performance of optical designs at both microwave and 
visible wavelengths. Based on this review we find a simple 
relationship between the (antenna-based) aperture efficiency 
and the Strehl ratio. We also show the results of several tests
performed to check the validity of this relationship that we carried out
using a ray-tracing software, ZEMAX and a full Physical Optics
software, GRASP9.3, 
applied to three different telescope designs.  

The paper is organized as follows: In Sect.~\ref{sec:gain} we review 
and discuss the definitions of antenna gain and aperture efficiency
while in Sect.~\ref{sec:strehl} we analyse the definition of
Strehl ratio and derive a simple relationship between the 
aperture efficiency and the Strehl ratio; in Sect.~\ref{sec:test}
we show the results of a comparison obtained using a Physical Optics
and a ray-tracing program and, finally, we draw our conclusions in 
Sect.~\ref{sec:concl}.

\section{Antenna gain and aperture efficiency}
\label{sec:gain}

\subsection{Definitions}
\label{sec:gaindef}


The gain of an antenna is a measure of the coupling of the antenna to
a plane wave field, and it can be written in terms of the {\it effective
area} (we assume that ohmic losses are negligible):
\begin{equation}
G(\theta,\phi) = \frac{4 \pi}{\lambda^2} \, A_{eff}(\theta,\phi) \, .
\end{equation}
%

\begin{figure}[htbp]
\centering
\includegraphics[width=11cm,angle=270]{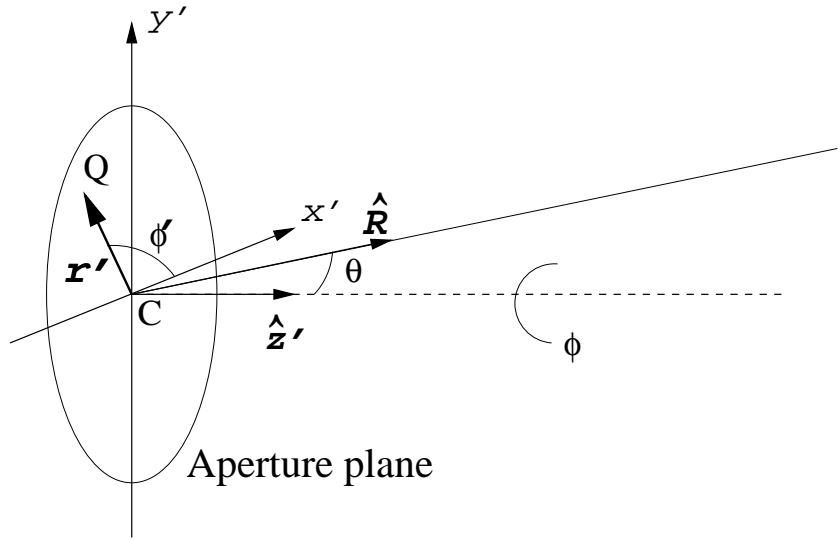}
\caption[ ]{
Coordinate systems used to calculate the antenna gain.
}
\label{fig:geomr}
\end{figure}

For an aperture type antenna the gain is expressible
in terms of the illumination by the feed. We can assume that the 
illumination is linearly polarized, and that the aperture lies on
an infinite plane. In this case the gain is expressible in terms of
$E_a(\bf r')$, the magnitude of the (in-phase) illuminating
electric field in the aperture plane. If almost all of the energy in the 
field is contained in a small angular region about the $z'$ axis, and if
we use the scalar-field approximation, then $G(\theta,\phi)$ can be
written as \cite{Silver}:
\begin{eqnarray}
G(\theta,\phi) & = & \frac{4 \pi}{\lambda^2}
\frac{\left| \, {\displaystyle \int\limits_{AP'}}
{\cal E}_{\rm a}({\bf r'},{\bf \hat R})
\, dS' \right|^2}
{\int\limits_{\infty} E_a^2({\bf r'}) \, dS'} \, ,
\label{eq:gain}
\end{eqnarray}
with
\begin{eqnarray}
{\cal E}_{\rm a}({\bf r'},{\bf \hat R}) & \equiv & E_{a}({\bf r'}) \,
e^{j\Phi({\bf r'})}  e^{j k \bf \hat R \cdot \bf r'} \label{eq:E_complex}  \\
\bf \hat{R} \cdot \bf r' & = & r' \sin \theta \, \cos (\phi-\phi') \nonumber \\
dS' & = & r' \, dr' \, d\phi' \nonumber
\end{eqnarray}
where we have introduced the complex electric field in the aperture,
${\cal E}_{\rm a}({\bf r'},{\bf \hat R})$. We have also indicated with
$k=2\pi/\lambda$ the wavenumber, and the field point Q at
position $\bf r'$ on the aperture plane (see Fig.~\ref{fig:geomr}) 
has polar coordinates $(r',\phi')$. 
$\bf \hat{R}$ is the unit vector along the direction to the observation 
point, with $\theta$ representing the angle formed by the direction to the
observation point and the optical axis and $\phi$ being the angle measured
in the plane of scan, i.e. perpendicular to the optical axis ($\bf \hat{z}'$),
as shown in Fig.~\ref{fig:geomr}.
The integral in the numerator is calculated over the antenna aperture,
whereas the integral in the denominator must extend over the entire plane
if there is any spillover illumination in the case of reflector antennas. 

The phase aberration function, $\Phi(\bf r')$,
 in Eq.~(\ref{eq:E_complex}) defines the phase at
point~$\bf r'$ in the aperture plane, which accounts for any change
in the optical path length resulting from the structural deformation of the
primary reflector, the displacements of the secondary reflector and the feed.
Thus, it is in $\Phi(\bf r')$ that one can take into account the positions
of different feed--horns in a FPA.

For aperture type antennas, the effective aperture can be related directly
to the antenna geometric area, $A_g$, by means of the {\it aperture efficiency},
$\eta_{_A}(\theta,\phi)$ (e.g., see Ref. \citeonline{Rudge}),
\begin{equation}
A_{eff}(\theta,\phi) = A_g \, \eta_{_A}(\theta,\phi) \, .
\end{equation}
Therefore,
\begin{eqnarray}
G(\theta,\phi) & = & \frac{4 \pi A_g }{\lambda^2} \,
\eta_{_A}(\theta,\phi) \\
\eta_{_A}(\theta,\phi) & = &
\frac{\left| \, {\displaystyle \int\limits_{AP'}}
{\cal E}_{\rm a}({\bf r'},{\bf \hat R})
\, dS' \right|^2}
{A_g \int\limits_{\infty} E_a^2({\bf r'}) \, dS'} \, .
\label{eta1}
\end{eqnarray}
The on-axis gain, $G_{\circ}$, is obtained by setting ${\bf \hat{R}} \cdot {\bf r'} = 0$,
then we obtain:
\begin{eqnarray}
G_{\circ} & = &  \frac{4 \pi A_g }{\lambda^2} \,
\eta_{\circ} \\
\eta_{\circ} & = &
\frac{\left| \, {\displaystyle \int\limits_{AP'}}
E_a({\bf r'}) \,
e^{j\Phi({\bf r'})} \, dS' \right|^2}
{A_g \int\limits_{\infty} E_a^2({\bf r'}) \, dS'} \, .
\label{etaon}
\end{eqnarray}
If the phase is constant over the aperture the on-axis gain attains its
maximum value, $G_{\rm M}$: 
\begin{eqnarray}
G_{\rm M} & = &  \frac{4 \pi A_g }{\lambda^2} \,
\eta_{\rm _M} \\
\eta_{\rm _M} & = &
\frac{\left| \, {\displaystyle \int\limits_{AP'}}
E_a({\bf r'}) \, dS' \right|^2}
{A_g \int\limits_{\infty} E_a^2({\bf r'}) \, dS'} \, .
\label{etamax}
\end{eqnarray}
A case of special interest is that of uniform illumination over the
aperture, i.e., $E_a({\bf r'})=const$ over the antenna aperture and zero outside. 
Hence, we obtain $\eta_{\rm _M}=1$ and
the {\it ideal gain}, $G_{\rm ideal}$, is then defined as
\begin{eqnarray}
G_{\rm ideal} & = & \frac{4 \pi A_g}{\lambda^2} \geq G_{\rm M}  \, .
\label{gideal}
\end{eqnarray}
Thus, we obtain the well-known result that the uniform field distribution
over the aperture gives the highest gain of all constant-phase
distributions over the aperture \cite{Silver}.

\subsection{Phase-error effects}
\label{sec:phase}

In the previous section we showed that if the
phase distribution is constant over the aperture, the maximum gain,
$G_{\rm M}$, is obtained in the direction of the optical axis,
i.e. ${\bf{\hat{R}}} \cdot {\bf{r'}} =$0.
%
\begin{figure}[htbp]
\centering
\includegraphics[width=12.5cm,angle=270]{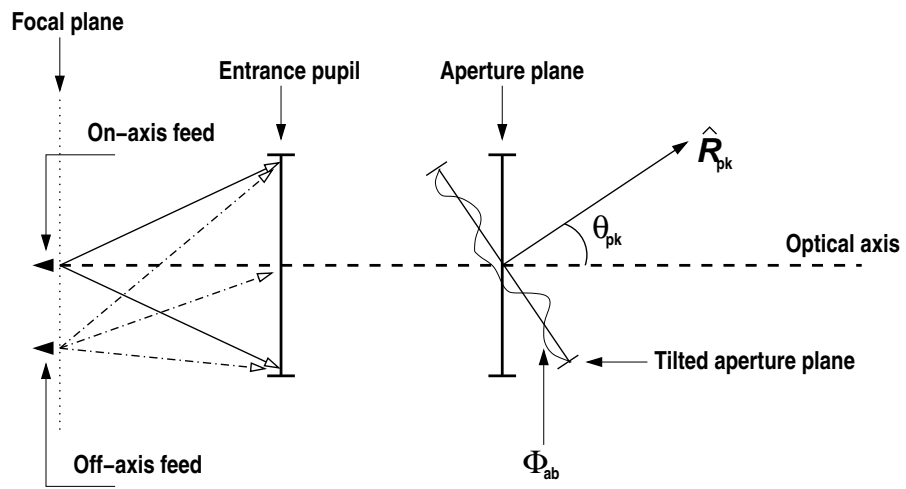}
\caption[ ]{
Off-axis feed and tilted aperture plane geometry.
}
\label{fig:tiltp}
\end{figure}
%
However, if a phase-error distribution is present over the aperture,
this may no longer be the case. A phase-error over the aperture, i.e.
deviations from uniform phase, may arise from various causes, such as
a displacement of the feed-horn from the on-axis focus (e.g., in FPAs),
or distortion of the optical surfaces, or it may be caused by phase-error
in the field of the feed-horn.

If the phase distribution is a {\it linear} function of the aperture
coordinates, then it can be shown that the far-field is the same as that
of the constant-phase distribution but displaced with respect to the
$z'-$axis, i.e. the direction of peak-gain is no longer in the direction
of the system optical axis \cite{Silver}.
In the case of arbitrary phase ditributions over the aperture, if the
phase-error does not deviate too widely from constant phase over the
aperture, and if it can be decomposed into a linear phase distribution
and higher-order terms,  then we may write
\begin{eqnarray}
\Phi({\bf r'}) &  = & \Phi_{\rm 1}({\bf r'}) + \Phi_{\rm ab}({\bf r'})
\label{phidef}
\end{eqnarray}
where $\Phi_{\rm 1}({\bf r'})$ is linear in the coordinates over the
aperture and causes an undistorted beam shift, i.e. a change in direction
of the peak gain (now corresponding to $\theta = \theta_{\rm pk}$), whereas
$\Phi_{\rm ab}({\bf r'})$ accounts for the true wave front distortion.
The shifted far-field beam can then be considered to have arisen
from a tilted aperture plane, i.e., from the aperture projected onto a
plane normal to the direction of the peak gain, ${\bf \hat{R}_{\rm pk}}$, as shown in
Fig.~\ref{fig:tiltp}.
In the projected aperture the linear phase distribution term cancels out,
leaving only higher-order phase errors, i.e.,
\begin{eqnarray}
\eta_{_A}(\theta,\phi) & = &
\frac{\left| \, {\displaystyle \int\limits_{AP}}
{\cal E}_{\rm a}({\bf r},{\bf \hat R})
\, dS \right|^2}
{A_g \int\limits_{\infty} E_a^2({\bf r}) \, dS} \, ,
\label{etaoff}
\end{eqnarray}
where now
\begin{equation}
{\cal E}_{\rm a}({\bf r},{\bf \hat R})  \equiv  E_{a}({\bf r}) \,
e^{j\Phi_{\rm ab}({\bf r})}  e^{j k \bf \hat R \cdot \bf r} \, ,
\end{equation}
and where $\bf r$ is the position of a point in the projected aperture plane,
indicated with $AP$, such that ${\bf {\hat{R}}} \cdot {\bf {r}} = 0$ 
for ${\bf{\hat{R}}} = \bf \hat{R}_{\rm pk}$.
If $\Phi_{\rm ab}({\bf r})=0$ then the field distribution has constant
phase over the projected aperture and the antenna gain in this aperture
will be given by \cite{Silver},
\begin{equation}
G_{\rm MP} = G_{\rm M} \, \cos \theta_{\rm pk}
\label{projgain}
\end{equation}
where $\cos \theta_{\rm pk} \simeq 1$ for most radio astronomical applications.
Therefore, in the following sections we will
refer to the antenna gain and aperture efficiency as the gain and
aperture efficiency in the projected aperture plane, unless noted 
otherwise.

\subsection{Main contributions to the aperture efficiency}
\label{sec:effic}


The aperture efficiency of an antenna is determined by a number of
phenomena and hence it can be written as the product of a number of
individual contributions (e.g., see Ref. \citeonline{Balanis}):
\begin{equation}
\eta_{_A}(\theta,\phi) = \eta_{\rm spill} \, \eta_{\rm taper}(\theta,\phi) \, \eta_{\rm phase}(\theta,\phi)
\label{eq:etagen}
\end{equation}
where $\eta_{\rm spill}$ is the {\it spillover efficiency}, $\eta_{\rm taper}$ is the
{\it taper efficiency} and $\eta_{\rm phase}$ takes into account
all {\it phase-error effects} causing a distortion of the
wave front. We have also assumed that ohmic losses are negligible 
and that the aperture is unblocked.
The spillover efficiency includes all
spillover contributions from the feed, subreflector, diffraction, etc.,
\begin{eqnarray}
\eta_{\rm spill}  =
\frac{ {\displaystyle \int\limits_{AP}}
E_a^2({\bf r}) \, dS }
{ {\displaystyle \int\limits_{\infty}}
 E_a^2({\bf r})  \, dS } \, .
\label{etas}
\end{eqnarray}
$\eta_{\rm taper}$ accounts for the aperture illumination taper due to the feed and
the reflector geometry,
\begin{eqnarray}
\eta_{\rm taper}(\theta,\phi) =
\frac{\left| \, {\displaystyle \int\limits_{AP}}
E_a({\bf r}) \, e^{j k \bf \hat R \cdot \bf r} \, dS \right|^2}
{A_g \int\limits_{AP} E_a^2({\bf r}) \, dS} \, ,
\label{etat}
\end{eqnarray}
and finally, $\eta_{\rm phase}$ accounts for the residual high-order phase
distortions of the wave-front at the aperture plane, due to
optical aberrations, surface errors or misalignments, etc.,
\begin{eqnarray}
\eta_{\rm phase}(\theta,\phi) =
\frac{\left| \, {\displaystyle \int\limits_{AP}}
{\cal E}_{\rm a}({\bf r},{\bf \hat R})
\, dS \right|^2}
{\left| \, {\displaystyle \int\limits_{AP}}
E_a({\bf r}) \, e^{j k \bf \hat R \cdot \bf r}
\, dS \right|^2} \, .
\label{etaab}
\end{eqnarray}
In the direction of the peak gain ${\bf{\hat R}} \cdot {\bf {r}} = 0$, 
as we earlier 
mentioned, and thus the $\bf{\hat{R}}=(\theta,\phi)$ dependence
can be dropped from $\eta_{\rm taper}$ and $\eta_{\rm phase}$. 

In the case of on-axis, dual-reflector systems the central subreflector
and its support structure cause a partial shadowing of the aperture, which
leads to a loss of efficiency. To take this effect into account the
integral at the numerator of Eq.~(\ref{etaoff}) can be written in the
case of a partially blocked aperture:
\begin{eqnarray}
& &  {\displaystyle \int\limits_{AP_{\rm block}}}
{\cal E}_{\rm a}({\bf r},{\bf \hat R})
\, dS  = 
%
{\displaystyle \int\limits_{AP}}
{\cal E}_{\rm a}({\bf r},{\bf \hat R})
\, dS  
%
- {\displaystyle \int\limits_{subr}}
{\cal E}_{\rm a}({\bf r},{\bf \hat R})
\, dS
\label{eq:integrals}
\end{eqnarray}
where $AP_{\rm block}$ represents the area of the aperture plane subtracted
of the blocked part, $AP$ indicates as usual the full area of the
aperture plane and $subr$ indicates the integration area
over the subreflector, assuming this is the main source of blockage. 
By substituting Eq.~(\ref{eq:integrals}) into
Eq.~(\ref{etaoff}) we thus obtain,
\begin{eqnarray}
\eta_{_A}(\theta,\phi) & = &
\frac{\left| \, {\displaystyle \int\limits_{AP}}
{\cal E}_{\rm a}({\bf r},{\bf \hat R})
\, dS \right|^2}
{A_g \int\limits_{\infty} E_a^2({\bf r}) \, dS} \times \nonumber \\
& & \left |  1 -
\frac{ {\displaystyle \int\limits_{A_{\rm subr}}}
{\cal E}_{\rm a}({\bf r},{\bf \hat R})
\, dS }
{ {\displaystyle \int\limits_{AP}}
{\cal E}_{\rm a}({\bf r},{\bf \hat R})
\, dS }
\right|^2
\label{etafull}
\end{eqnarray}
where the first term at the right can once again be written as in
Eq.~(\ref{eq:etagen}) and thus the second term can be interpreted as the
{\it blocking efficiency} due to the subreflector,
\begin{eqnarray}
\eta_{\rm block}(\theta,\phi) = \left |  1 -
\frac{ {\displaystyle \int\limits_{A_{\rm subr}}}
{\cal E}_{\rm a}({\bf r},{\bf \hat R})
\, dS }
{ {\displaystyle \int\limits_{AP}}
{\cal E}_{\rm a}({\bf r},{\bf \hat R})
\, dS }
\right|^2 \, .
\label{etablock}
\end{eqnarray}
We note that in the direction of the peak-gain 
(${\bf{\hat R_{\rm pk}}} \cdot {\bf{r}} = 0$),
for an uniform, unaberrated
($\Phi_{\rm ab}({\bf r})=0$) field we find the well-known result,
\begin{equation}
\eta_{\rm block} = \left ( 1 - \frac{A_{\rm subr}} {A_{\rm prim}} \right )^2
\label{etablockdue}
\end{equation}
where $A_{\rm prim}$ and $A_{\rm subr}$ are the surface areas of the
primary and secondary reflectors, respectively. In general, the 
geometrical blockage caused by the support struts can be up to several times
larger than the blockage caused by the secondary mirror, especially in
open-air antennas. Therefore, the blockage efficiency given by 
Eq.~(\ref{etablockdue}) usually overestimates the real efficiency and should
be corrected including the strip blockage of the plane-wave and the 
blockage from the converging spherical-wave between the primary mirror and 
the subreflector (e.g., see Ref. \citeonline{Lamb}).


\section{Strehl ratio}
\label{sec:strehl}

\subsection{Strehl ratio on-axis}
\label{sec:strehlon}


\begin{figure}[htbp]
\centering
\includegraphics[width=9cm,angle=270]{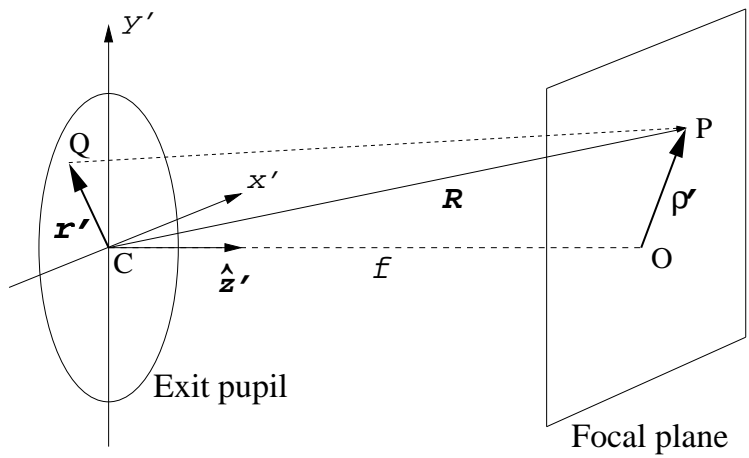}

\caption[ ]{
Coordinate frame at exit pupil $(x',y',z')$ and position, 
$\mbox{\boldmath $\rho'$}$, of point $P$ at focal plane.
The field amplitude at point $\mbox{\boldmath $r'$}$ on the
system's exit pupil is $E_{ex}({\bf r'})$.
}
\label{fig:geomo}
\end{figure}

While the main antenna-based figures-of-merit are usually, though not
necessarily, defined in the far field of the aperture, the Strehl
ratio of an optical imaging system is defined as the ratio of the
aberrated to unaberrated incoherent Point Spread Function (PSF,
hereafter \cite{Mahajan}).

When considering the optical system in receiving mode, the PSF refers 
to the instantaneous field distribution in the focal
plane of an optical imaging system produced by a far-field point source.
For simplicity we assume that
the fields are emerging from the exit pupil of the optical system with
a system focal length $f$, and converging towards the image plane.
Let's suppose that the exit pupil is on an infinite plane located at
$z=0$, and with the normal unit vector in direction of the z-axis,
$\bf \hat n =\bf \hat z'$ (see Fig.~\ref{fig:geomo})\footnote{The focal plane and the 
observation point in the far-field defined by $\bf \hat R$ in the previous
sections (where the optical system was considered in
transmission mode) lay on opposite
directions with respect to the $x'y'$ plane. This will be taken into
account in Sect.~\ref{sec:strehlgain}}. 
Then, following Ref. \citeonline{Mahajan},\citeonline{Schroeder}, 
the scalar field at a point $P$ at position ${\mbox{\boldmath $\rho'$}}$
in the paraxial focal plane (see Fig.~\ref{fig:geomo}) is given by
\begin{eqnarray}
E_f(\mbox{\boldmath $\rho'$}) & \propto & \,
{\int\limits_{AP'}}
\vphantom{\frac{k}{Z_o}}
E_{ex}({\bf r'}) \, e^{- j \frac{k}{f} \mbox{\boldmath \scriptsize $\rho'$} \cdot \bf r'} \, dS' 
\label{eq:foc}
\end{eqnarray}
%
where $E_{ex}$ is the field amplitude at
a point $Q$ at position $\bf r'$ on the system's exit pupil 
and $f$ is also equal to the radius of curvature of the reference
sphere centered at point $O$ in the focal plane.
In the case of a point source in the far field of the system
$E_{ex}$ is uniform over the pupil.

In Eq.~(\ref{eq:foc}) the substitution of the exit pupil
for the antenna aperture plane, and the consequent use of $\bf r'$ in both
cases, is justified by using the {\it equivalent parabola} (e.g., in a
dual-reflector system) and by the fact that
when the point source object is at infinity, then the diameter (assuming
a circular aperture) of the exit pupil can be substituted with the diameter of
the entrance pupil, or main dish in a dual-reflector system
(see Ref. \citeonline{Schroeder}, p. 184), and the system focal length would be in
this case the focal length of the equivalent parabola \cite{Hannan}.
In other words, the spherical (i.e., aberration-free) wavefront
leaving the equivalent parabola and converging to the focus is identified
here with the Gaussian reference sphere centred on the exit pupil.
Then, we can state that {\it the (unaberrated) incoherent PSF is 
simply the square modulus of $E_f(\mbox{\boldmath $\rho'$})$, i.e.,
$ {\rm PSF} = I({\mbox{\boldmath $\rho'$}}) = 
|E_f({\mbox{\boldmath $\rho'$}})|^2$.
}

Eq.~(\ref{eq:foc}) is strictly valid in the absence of phase errors
that may modify the perfectly spherical
convergent wave that was assumed earlier in the special case
of an aberration-free wave-front.
In the more general case of a distorted wave-front Eq.~(\ref{eq:foc}) should
be re-written as:
\begin{eqnarray}
E_f({\mbox{\boldmath $\rho'$}}) & \propto & \,
{\int\limits_{AP'}}
\vphantom{\frac{k}{Z_o}}
E_{ex}({\bf r'}) \, e^{- j \frac{k}{f} {\mbox{\boldmath \scriptsize $\rho'$}} \cdot \bf r'} \,
e^{j\Phi(\bf r')} \, dS'
\label{eq:focdis}
\end{eqnarray}
where $\Phi(\bf r')$ is the phase error term.
The Strehl ratio, $S$, of the imaging system is then given by the ratio
of the {\it central} (i.e., ${\mbox{\boldmath $\rho'$}}=0$)
irradiance of its aberrated and unaberrated PSFs.
From Eq.~(\ref{eq:focdis}) $S$ can be written in the 
form {\cite{Mahajan}$^{,}$\cite{Born}}:
\begin{eqnarray}
S_{\circ}  =   \frac{I(0)}
{I(0)|_{\Phi=0}} =
\frac{\left| \int\limits_{AP'} E_{ex}({\bf r'}) \, e^{j\Phi(\bf r')} \, dS' \right|^2}
{\left| \int\limits_{AP'} E_{ex}({\bf r'}) \, dS' \right|^2}
\label{eq:StrehlRatioStd1}
\end{eqnarray}
where $S_{\circ} \equiv S({\mbox{\boldmath $\rho'$}}=0)$.
The Strehl ratio can also be used as a measure of the on-axis PSF away from
its central irradiance peak, and thus we can write:
\begin{eqnarray}
S({\mbox{\boldmath $\rho'$}}) & = &  \frac{I({\mbox{\boldmath $\rho'$}})}
{I(0)|_{\Phi=0}} = \nonumber \\
& = & \frac{\left| \int\limits_{AP'} E_{ex}({\bf r'}) \,
e^{- j \frac{k}{f} \mbox{\boldmath \scriptsize $\rho'$} \cdot \bf r'} \,
e^{j\Phi(\bf r')} \, dS' \right|^2}
{\left| \int\limits_{AP'} E_{ex}({\bf r'}) \, dS' \right|^2} \, .
\label{eq:StrehlRatioStd2}
\end{eqnarray}
%

\subsection{Strehl ratio off-axis}
\label{sec:strehloff}

In equations~(\ref{eq:foc}) to (\ref{eq:StrehlRatioStd2}) the position
in the paraxial focal plane of the central irradiance peak of the PSF
was taken as the origin of a Cartesian system of axes and also as
the center of the (unaberrated) Gaussian reference sphere \cite{Born}.
The observation of an object point off-axis, which is equivalent to having
the feed lateraly displaced in a microwave antenna, introduces both a
change in the position of the PSF peak (or direction of peak gain in
an antenna) and wave-front aberration.
The quasi-spherical (i.e., aberrated) wave will be thus converging to a
point displaced with respect to point $O$ in Fig.~\ref{fig:geomo}. If
$\mbox{\boldmath{$\rho'$}}_{\rm pk}$ represents the position of the off-axis PSF peak in the
focal plane, then Eq.~(\ref{eq:StrehlRatioStd2}) can be re-written as:

%
\begin{eqnarray}
S(\mbox{\boldmath $\rho'$}) &  = &  \frac{I(\mbox{\boldmath $\rho'$})}
{I(\mbox{\boldmath $\rho'$}_{\rm pk})|_{\Phi_{\rm ab}=0}} = \nonumber \\
& = & \frac{\left| \int\limits_{AP'} E_{ex}({\bf {r'}}) \,
e^{- j \frac{k}{f} \mbox{\boldmath \scriptsize $\rho'$} \cdot \bf r'} \,
e^{j\Phi(\bf r')} \, dS' \right|^2}
{\left| \int\limits_{AP'} E_{ex}({\bf {r'}}) \,
e^{- j \frac{k}{f} \mbox{\boldmath \scriptsize $\rho'$}_{\rm pk} \cdot \bf r'} \,
e^{j\Phi_1(\bf r')} \, dS' \right|^2}
\label{eq:StrehlRatioStd3}
\end{eqnarray}
where $\Phi_{\rm ab}$ and $\Phi_1(\bf r')$ have been defined in 
Eq.~(\ref{phidef}) and Sect.~\ref{sec:phase}. 
Thus, $I(\mbox{\boldmath $\rho'$}_{\rm pk})|_{\Phi_{\rm ab}=0}$
represents the peak irradiance of the {\it un}aberrated, off-axis PSF.

In Sect.~\ref{sec:phase} we saw that by tilting the aperture plane
so that it becomes perpendicular to the direction of the peak gain, it is
possible to write the aperture efficiency in terms of $\Phi_{\rm ab}$
only. Likewise, in the definition of the PSF it is possible to
align the $z$-axis along the direction from the center of the exit pupil
to the off-axis Gaussian image point, which can also be taken as the
origin of a new Cartesian system of axes. The Gaussian image point is also
the center of curvature of the (tilted) wave front, and for this point
all path lengths from the spherical wave front would be equal, in the
absence of higher-order phase distortions.
Then, Eq.~(\ref{eq:StrehlRatioStd3})
takes the same form as Eq.~(\ref{eq:StrehlRatioStd2}), i.e.
\begin{eqnarray}
S(\mbox{\boldmath $\rho$}) &  =  &  \frac{I(\mbox{\boldmath $\rho$})}
{I(0)|_{\Phi_{\rm ab}=0}} = \nonumber \\
& = & \frac{\left| \int\limits_{AP} E_{ex}({\bf r}) \,
e^{- j \frac{k}{f} \mbox{\boldmath \scriptsize $\rho$} \cdot \bf r} \,
e^{j\Phi_{\rm ab}(\bf r)} \, dS \right|^2}
{\left| \int\limits_{AP} E_{ex}({\bf r}) \, dS \right|^2}
\label{eq:StrehlRatioStd4}
\end{eqnarray}
where the peak of the PSF is now at point $\mbox{\boldmath $\rho$}=0$ in the
new system of axes, centered on the Gaussian image point in the
focal plane, and
$\bf r$ now lies on a tilted plane, $AP$, perpendicular to the direction
of the off-axis PSF peak. Thus we have in the projected plane,
\begin{eqnarray}
S_{\circ} &  =  &  \frac{I(0)}
{I(0)|_{\Phi_{\rm ab}=0}} = \nonumber \\
& = & \frac{\left| \int\limits_{AP} E_{ex}({\bf r}) \,
e^{j\Phi_{\rm ab}(\bf r)} \, dS \right|^2}
{\left| \int\limits_{AP} E_{ex}({\bf r}) \, dS \right|^2} \, .
\label{eq:StrehlRatioStd5}
\end{eqnarray}
%

\subsection{Strehl ratio and aperture efficiency}
\label{sec:strehlgain}


In this section we use the previous results to derive a relationship
between aperture efficiency and Strehl ratio. First,
we use Eq.~(\ref{etaoff}) to form the ratio of the aberrated and
unaberrated aperture efficiency (in the projected aperture plane), i.e.
\begin{eqnarray}
\frac{\eta_{_A}(\theta,\phi)}
{\eta_{_{\rm MP}}}  & = & 
%
\frac{\left| \, {\displaystyle \int\limits_{AP}}
{\cal E}_{\rm a}({\bf r},{\bf \hat R}) \, dS \right|^2} 
{\left| \, {\displaystyle \int\limits_{AP}}
E_a({\bf r})
\, dS \right|^2} \, ,
\label{etaratio}
\end{eqnarray}
with
\begin{eqnarray}
\eta_{_{\rm MP}} & = & 
\frac{\left| \, {\displaystyle \int\limits_{AP}}
E_a({\bf r}) \, dS \right|^2}
{A_g \int\limits_{\infty} E_a^2({\bf r}) \, dS}
\label{etamp}
\end{eqnarray}
where $\eta_{_A}({\bf \hat R}) \equiv \eta_{_A}(\theta,\phi)$ is the
aberrated aperture efficiency measured in the generic direction 
$\bf \hat R=(\theta,\phi)$ (i.e., not coincident with 
the direction of the
peak gain, ${\bf \hat{R}_{\rm pk}}$), for the general case in which the
direction of peak-gain is not along the main optical axis of the system,
as explained in Sect.~\ref{sec:phase}. 
$\eta_{_{\rm MP}} \equiv \eta_{_A}({\bf \hat{R}_{\rm pk}})|_{\Phi_{\rm ab}=0}$
is the unaberrated aperture efficiency measured in the direction of the
(off-axis) peak gain, i.e. $\eta_{\rm MP}$ represents the peak aperture efficiency
as measured in the projected aperture plane. Recalling that in the direction 
of the peak-gain ${\bf \hat{R}_{\rm pk}} \cdot {\bf r} =0$ 
(see Sect.~\ref{sec:effic})
the $\bf \hat R$-dependence can be dropped from $\eta_{\rm MP}$.
From eqs.~(\ref{etamax}) and (\ref{projgain}) it also follows that,
\begin{equation}
\eta_{_{\rm MP}} = \eta_{_{\rm M}} \, \cos \theta_{\rm pk} \simeq \eta_{_{\rm M}}
\label{etaprojrelat}
\end{equation}
if $\theta_{\rm pk} << 1$, where $\eta_{_{\rm M}}$ is the maximum aperture efficiency as defined
in Sect.~\ref{sec:gaindef}. From Eq.~(\ref{etamp}) and equations~(\ref{etas}) 
and (\ref{etat}) we also see that $\eta_{_{\rm M}} = \eta_{\rm spill} \, \eta_{\rm taper}$.

Then, we note that equations~(\ref{eq:StrehlRatioStd4}) and (\ref{etaratio})
have the same form and, for small angles close to the optical axis it holds 
that
\begin{eqnarray*}
\mbox{\boldmath $\alpha$} \cdot {\bf r} = - {\bf \hat R} \cdot \bf r 
\end{eqnarray*}
where we have defined $\mbox{\boldmath $\alpha$} = \mbox{\boldmath $\rho$} / f$ (see the discussion in
Ref. \citeonline{Padman}).
However, since $E_{\rm ex}$ represents the field produced by a point source
in the far field of the system, in order to conclude that
equations~(\ref{eq:StrehlRatioStd4}) and (\ref{etaratio}) are fully
equivalent one must assume that the incident field on the optical
system from a distant source has an apodization equivalent to that
produced by the feed illumination on the antenna aperture 
(see Sect.~\ref{sec:strehlon}).
In this case we can write $E_{\rm ex}({\bf r}) = E_{\rm a}({\bf r}) $, and thus
\begin{equation}
\eta_{_A}({\bf \hat R}) = {\eta_{_{\rm M}}} \, S(\mbox{\boldmath $\rho$}) \, .
\label{eq:seta}
\end{equation}
Then, by comparing equations~(\ref{eq:etagen}) to (\ref{etaab}) with
Eq.~(\ref{eq:seta}) one can see that in general,
\begin{equation}
\eta_{_{\rm M}} \, S(\mbox{\boldmath $\rho$}) = \eta_{\rm spill} \, \eta_{\rm taper}(\bf \hat R) \, \eta_{\rm phase}(\bf \hat R) \, .
\end{equation}
Usually, however, one is interested in the aperture efficiency at the
nominal position of the peak gain (i.e., at the center of the far-field beam),
or equivalently at the center of the PSF, then it also holds that
%
%
%
\begin{equation}
\left \{ \begin{array}{l}
\eta_{\circ} = {\eta_{\rm _M}} \, S_{\circ} \\
{\eta_{_{\rm M}}} = \eta_{\rm spill} \, \eta_{\rm taper} \\
\end{array}
\right.
\label{eq:setapk}
\end{equation}
%
and
\begin{equation}
S_{\circ} = \eta_{\rm phase}
\label{eq:sph}
\end{equation}
with ${\bf \hat{R}_{\rm pk}} \cdot {\bf r} = 0$ and
$\eta_{\circ}=\eta_{_A}({\bf \hat R}={\bf \hat{R}_{\rm pk}})$ is the aperture efficiency in the
direction of the peak-gain, corresponding to Eq.~(\ref{etaon})
in the projected aperture plane, i.e.
\begin{eqnarray}
\eta_{\circ} =
\frac{\left| \, {\displaystyle \int\limits_{AP}}
E_a({\bf r}) \,
e^{j\Phi_{\rm ab}({\bf r})} \, dS \right|^2}
{A_g \int\limits_{\infty} E_a^2({\bf r}) \, dS}
\label{etaonproj}
\end{eqnarray}
where we have not used the subscript ``$p$'' (for ``projected parameter'')
in ${\eta_{\circ}}$ because of the approximation in Eq.~(\ref{etaprojrelat}). 
Therefore, {\it Eq.~(\ref{eq:sph}) finally shows 
the equivalence between the Strehl ratio and phase efficiency}.

Clearly, $\eta_{_{\rm M}}$ takes into account both taper and
spillover effects, whereas $S_{\circ}$ is a measure of the phase aberrations.
Therefore, in the case of an {\it un}aberrated wave front, i.e. 
$S_{\circ}  =  \eta_{\rm phase} = 1$, the aperture efficiency is ${\eta_{\circ}}  =  \eta_{_{\rm M}}$ and
depends only on the spatial distribution of the field over the antenna
aperture.
Furthermore, by explicitly writing the aberration function, 
$\Phi_{\rm ab}({\bf r})$, in terms of the primary aberrations 
(e.g., see Ref. \citeonline{Mahajan}) it would
be possible to derive the individual contributions to the aperture 
efficiency by, e.g., coma, astigmatism and curvature of field, which 
are usually the most relevant aberrations in radiotelescopes.
However, this is beyond the scopes of this work and will not be done
here.

\section{Comparison of Strehl ratio and  aperture efficiency }
\label{sec:test}

In this section we want to compare the values of the Strehl 
ratio, obtained from a ray-tracing optical software, ZEMAX~(Focus 
Software \cite{zemax}), and the associated value of $\eta_{\rm phase}$, obtained 
through the numerical integration of Eq.~(\ref{etaab}) and using the 
aperture field values computed by a Physical Optics program,
GRASP9.3 ~(TICRA Engineering Consultants \cite{grasp}).  Several configurations
have been analysed and are discussed below.



\subsection{Description of software packages}
\label{sec:soft}

The analysis has been conducted using the GRASP9.3 package,
which is a commercial tool for calculating the electromagnetic
radiation from systems consisting of multiple reflectors with several
feeds and feed arrays. This package can use several high-frequency
techniques for the analysis of large reflector antennas, such as Physical
Optics (PO) supplemented with the Physical Theory of Diffraction (PTD),
Geometrical Optics (GO) and Uniform Geometrical Theory of Diffraction (GTD),
which require a moderate computational effort.

The PO technique is an accurate method that
gives an approximation to the surface currents valid for perfectly
conducting scatterers which are large in terms of wavelengths. The PO
approximation assumes that the current in a specific point on a curved
but perfectly conducting scatterer is the same as the current on an infinite
planar surface, tangent to the scattering surface. For a curved surface,
the PO current is a good approximation to the actual one if the dimensions of
the scattering surface and its radius of curvature are sufficiently large
measured in wavelengths.
The well-known GO method uses ray-tracing techniques for describing wave
propagation. Since GO gives discontinuities in the total electromagnetic
field, GTD is often applied in addition to GO, since GTD methods may account
for diffraction effects.


On the other hand, ZEMAX is a classical optical design tool based on
ray-tracing methods, which combines three major categories of analysis in
one package: lens design, physical optics, and non-sequential
illumination/stray light analysis.

\subsection{Calculation of the aperture efficiency with GRASP9.3}
\label{sec:srtgain}

As described in Sect.~\ref{sec:soft}, GRASP9.3 allows several methods for the
electromagnetic analysis of the reflecting surfaces. 
An interesting tool of GRASP9.3, based on the ray-tracing, 
for calculating the aperture
field is the so-called ``Surface Grid'' \cite{ref_grasp}. This method returns the
reflected magnetic field on the surface according to the formula:
${\bf {H_{\rm r}}}={\bf {H_{\rm i}}}-2{\bf \hat{n}}({\bf \hat{n}} \cdot {\bf {H_{\rm i}}})$, 
where $\bf H_{\rm i}$ is the magnetic incident field 
and $\bf \hat{n}$ is the normal to the surface. The magnetic reflected field on 
the surface, $\bf H_{\rm r}$, is then
projected, with a phase adjustment, on the aperture plane. As described in
Sect.~\ref{sec:phase}, when the feed is placed off-axis the aperture 
plane is tilted according to the direction of the peak-gain. For a dual 
reflector configuration, the scattering from the secondary and 
primary mirrors has been analyzed through the GTD technique and
the ``Surface Grid'', respectively.

This approach is particularly appropriate when the diameter-to-wavelength 
ratio of the primary reflector is very large and when the observation point 
is in the near-field (such as the aperture plane case). Under these 
conditions the PO method would be very time-consuming; in fact, it would 
require a huge number of points on the reflector where currents need to be 
evaluated. Using the method described here to analyze the primary 
reflector the diffracted field from the edge of the reflector is not 
considered.  However, the numerical results obtained with this ``hybrid'' 
technique have been compared with those obtained by applying
the PO method to both the primary and secondary mirrors, resulting
in a very good agreement between the two methods.

In order to calculate the aperture efficiency from Eq.~(\ref{etaratio})
we use the the complex electric field in the aperture plane,
i.e. ${\cal E}_{\rm a}(\bf r,\bf \hat R)$, 
produced by GRASP9.3, which is
tabulated through its real and imaginary components. These can then be 
used to calculate the amplitude and the phase function of the field. 
%
%
%
%
The complex electric field is finally read by a proprietary code 
which evaluates Eq.~(\ref{etaab}) in order to determine the phase efficiency.

\subsection{Comparison of results}
\label{sec:res}

The values of the Strehl ratio and phase efficiency obtained with ZEMAX
and GRASP9.3, respectively, have been compared using three different optical
systems. These systems have been selected to
represent standard telescope designs, and the frequencies used in the 
simulations cover the mm- and submm-wavelength regimes.
For the electromagnetic analysis with GRASP9.3, we have always used
a linearly polarized Gaussian feed. 
Although more realistic feed models to describe circular horns could 
be adopted, for the sake of comparison with ZEMAX and to avoid 
introducing any systematic error due to different feed illumination,
we report the results obtained with a Gaussian
model only. The level of apodization in ZEMAX has then been chosen to be
consistent with that produced by the Gaussian feed-horn.

\subsubsection{Single-dish antenna}
\label{sec:single}


First, we have carried out the comparison in the simplest possible case, 
i.e. an unblocked spherical reflector antenna.
This choice eliminates or minimises potential 
discrepancies due to different handling in ZEMAX and GRASP9.3 of effects such 
as multiple reflections, aperture blocking and diffraction at secondary 
surfaces.

\begin{figure}[htbp]
\centering
\includegraphics[width=10cm]{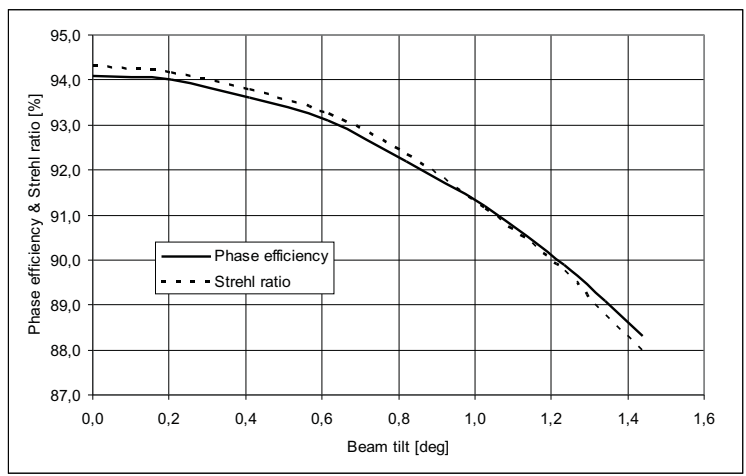}
%
%

\caption[ ]{
Plot of the Strehl ratio and of the phase efficiency at a
wavelength of 500$\mu$m for the
case of a spherical reflector 105 cm in diameter with a $f/\#=2$.
}
\label{fig:sph}
\end{figure}

The surface chosen for this simulation is spherical because it ensures that
spherical aberration will limit the overall FOV to small ($\la 1^\circ$) angles 
near the optical axis. This is required in order to avoid introducing further
variables in the comparison between ZEMAX and GRASP9.3 due to the incidence 
angle of radiation over the aperture of the feed-horn in the focal plane, 
which may affect the coupling between the PSF and the electric fields on 
the horn aperture. The selected aperture was 105 cm in diameter 
with a $f/\#=2$ and the simulations have been carried out at a wavelength of 
500 $\mu$m. 
For the electromagnetic analysis with GRASP9.3, a linearly polarized Gaussian 
feed has been used with a taper level of $-12$~dB at $14^\circ$.

The results are shown in Fig.~\ref{fig:sph}: the comparison has been extended
up to a maximum offset angle of $\simeq 1.4^\circ$, or about 44 beams at 
500~$\mu$m, and the maximum measured difference between the Strehl ratio 
calculated by ZEMAX and the phase efficiency calculated by GRASP9.3 is 0.38\% 
at the maximum offset angle. We also note, however, a 0.25\% discrepancy on boresight, which will be discussed in the next section.

\subsubsection{Dual-reflector antenna: Cassegrain configuration}
\label{sec:cass}

%
%

We have then analysed the most common radio telescope design, consisting of 
a dual-reflector antenna. We first consider the classical Cassegrain 
configuration, which we have derived from the design of the 
``Balloon-borne Large Aperture Submillimeter Telescope'' (BLAST) telescope \cite{dev04}. 
Compared to the original design with a spherical primary mirror \cite{olmi01} and to the newer telescope design with a 
Ritchey-Chretien optical 
configuration, the system analysed here has a parabolic primary and a 
hyperbolic secondary. The diameters of primary and secondary
mirrors are 181.61 and 42.76 cm, respectively, and the system focal ratio is 5.
As in the single-reflector case, a linearly polarized Gaussian feed has been 
used, but with a taper level of $-9$~dB at $6^\circ$.

\begin{figure}[htbp]
\centering
\includegraphics[width=10cm]{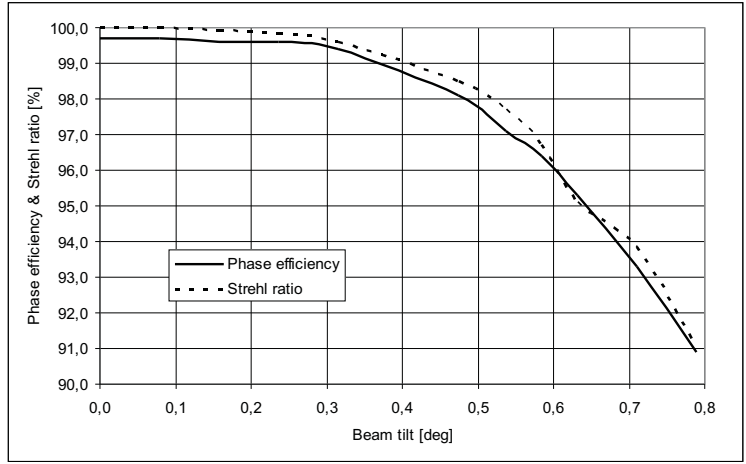}
\caption[ ]{
Plot of the Strehl ratio and of the phase efficiency at a
wavelength of 500$\mu$m for the
case of a classical Cassegrain telescope.
The diameters of primary and secondary mirrors are 181.61 and 42.76 cm,
respectively, and the system focal ratio is 5.
}
\label{fig:blast}
\end{figure}

\begin{figure}[htbp]
\centering
\includegraphics[width=10cm]{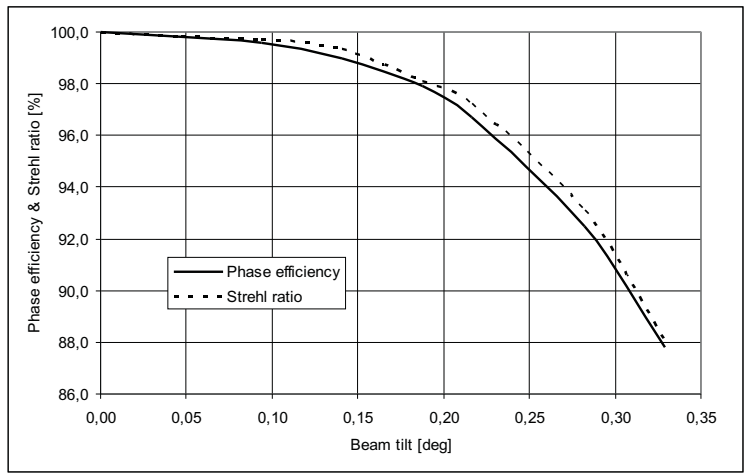}
\caption[ ]{
Same as Fig.~\ref{fig:blast} for the scaled-up version of the
BLAST telescope. The primary and secondary
reflector diameters equal to 12.2m and 2.6m, respectively.
}
\label{fig:blastesp}
\end{figure}

The results are shown in Fig.~\ref{fig:blast}: the comparison has been 
extended up to a maximum offset angle of $\simeq 0.79^\circ$, or about 42 beams 
at 500 $\mu$m, thus quite equivalent to the previous simulation.
The maximum measured difference between the Strehl ratio calculated by ZEMAX
and the phase efficiency calculated by GRASP9.3 is about 0.59\% at an offset 
angle of about $0.5^\circ$. We observe that the discrepancy between the two methods
is also relevant ($0.2-0.3$\%) for offset angles near boresight and it is possibly more
systematic in this case than in the single-reflector design analysed
in the previous section. 

This on-axis difference is likely due to the relatively small 
secondary diameter to wavelength ratio, $D_{\rm sec}/\lambda$, which may 
cause an on-axis decrease of the antenna gain due to diffraction effects
from the edge of the secondary. To test this hypothesis, we have scaled-up
the BLAST telescope, while keeping constant the wavelength, in order 
to obtain an optical design with a much larger $D_{\rm sec}/\lambda$ ratio,
comparable to that used in the next section for the 
``Sardinia Radio Telescope''. We have thus obtained a telescope with the
same focal ratio at the Cassegrain focus but with a primary and secondary 
reflector diameter equal to 12.2m and 2.6m, respectively.

The results are shown in Fig.~\ref{fig:blastesp}: in this case the comparison has been 
extended up to a maximum offset angle of $\simeq 0.33^\circ$, or about 116 beams 
at 500 $\mu$m. As expected, the discrepancy near the optical axis has decreased
compared to both the single-dish and the original BLAST cases. The maximum
difference is about 0.61\%, thus still quite similar to that 
observed in the original BLAST design despite the much larger offset angle
in beam units used in the scaled-up telescope. These results indicate that
diffraction effects are calculated differently in GRASP9.3 and ZEMAX.

\subsubsection{Dual-reflector antenna: Gregorian configuration}
\label{sec:greg}


The third system analysed during this comparison is another dual-reflector 
antenna, though in  a Gregorian configuration. In this case we have 
changed the wavelength to a larger value of 3~mm and have also chosen a 
telescope with a much higher $D/\lambda$ ratio. The baseline design is in 
this case the ``Sardinia Radio Telescope'' (SRT \cite{Grueff}); 
however, we have 
converted the original {\it shaped} design of the SRT to a more standard Gregorian 
configuration, keeping the same aperture (64 m) and system focal ratio (2.34) 
of the SRT.
As in the previous two cases, a linearly polarized Gaussian feed has been
used, with a taper level of $-12$~dB at $12^\circ$.

\begin{figure}[htbp]
\centering
\includegraphics[width=10cm]{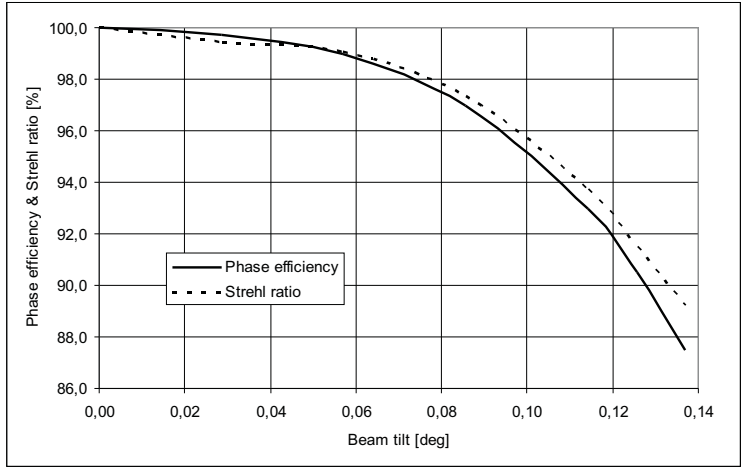}
\caption[ ]{
Plot of the Strehl ratio and of the phase efficiency at a
wavelength of 3~mm for the case of a classical
Gregorian  telescope, with a primary reflector diameter of
64~m and a system focal ratio of 2.34.
}
\label{fig:srt}
\end{figure}

The results are shown in Fig.~\ref{fig:srt}: the comparison has been extended
up to a maximum offset angle of $\simeq 0.136^\circ$, or about 42 beams at 
$\lambda = 3$~mm, thus consistent with the simulations used for the 
single-dish and the BLAST configurations.
The maximum measured difference between the Strehl ratio calculated by ZEMAX
and the phase efficiency calculated by GRASP9.3 is about 1.9\%, thus larger
than in the optical systems discussed above. However, in the range of
offset angles where the Strehl ratio (or equivalently the phase efficiency) 
is $> 0.95$, i.e. the range which is normally targeted by the optical
design of diffraction-limited telescopes, the difference between 
Strehl ratio and phase efficiency is $< 0.5$\%, consistent with 
that observed in the BLAST telescope.

%
%

\section{Conclusions}
\label{sec:concl}

We have reviewed the main design parameters generally used in evaluating the
performance of optical designs at both microwave and visible wavelengths.
In particular, we have reviewed the classical concept of antenna gain
and the main contributions to the aperture efficiency, with special
attention to phase-error effects. We have then described the 
formalism with which to compare the aperture efficiency and its components with
the Strehl ratio, which is the standard parameter used to evaluate
the image quality of diffraction-limited telescopes at visible/IR
wavelengths. We have shown that a simple relationship can be found
between Strehl ratio and aperture efficiency: the Strehl ratio is equal to the phase efficiency when the 
apodization factor is taken into account.

We have then compared these two parameters
by running ray-tracing software, ZEMAX and full Physical Optics 
software, GRASP9.3, on three different telescope designs: 
a single spherical reflector, a Cassegrain telescope and finally
a Gregorian telescope. These three configurations span a factor of 
$\simeq 10$ in terms of $D/\lambda$. The simple spherical reflector
allows the most direct comparison between Strehl ratio and
phase efficiency, as it is only marginally affected by edge diffraction
effects. In this case we find that these two parameters differ by
less than 0.4\% in our ZEMAX and GRASP9.3 simulations, up to an angle 
of about 44 beams off-axis. The other two 
configurations are more prone to diffraction effects caused by the
secondary reflector, especially in the case of the smaller Cassegrain
telescope.

The phase-efficiency is the most critical contribution to the 
aperture efficiency of the antenna, and the most difficult parameter to 
optimize during the telescope design. The equivalence between
the Strehl ratio and the phase efficiency gives the designer/user
of the telescope the opportunity to use the faster (and less
expensive) commercial ray-tracing software to optimize the
design using their built-in optimization routines.

\section*{Acknowledgments} 
This work was partly sponsored by the Puerto Rico NASA Space Grant
Consortium.

\end{document}